\documentclass[journal]{IEEEtran}
\usepackage{amsmath,amsfonts}
\usepackage{algorithmic}
\usepackage{algorithm}
\usepackage{array}
\usepackage[caption=false,font=normalsize,labelfont=sf,textfont=sf]{subfig}
\usepackage{textcomp}
\usepackage{stfloats}
\usepackage{url}
\usepackage{verbatim}
\usepackage{graphicx}
\usepackage{cite}
\begin{document}

\title{An Experimental Evaluation of Accurate Scheduling and Hardware Timestamping on NVIDIA ConnectX NICs}

\author{Takahiro Hirofuchi, Takaaki Fukai
\thanks{T. Hirofuchi and T. Fukai are with Intelligent Platform Research Institute (IPRI), National Institute of Advanced Industrial Sciences and Technology (AIST), Tokyo, Japan}%
}

\maketitle

\begin{center}
\textit{Preprint version. This manuscript has not been peer reviewed.
A revised version is planned for submission to a journal or an international conference.}
\end{center}

\begin{abstract}
High-precision packet transmission is becoming increasingly important in deterministic networking applications, including 5G fronthaul and Time-Sensitive Networking (TSN).
Recent NVIDIA ConnectX network interface cards (NICs) provide Accurate Scheduling, a hardware-assisted mechanism for transmitting Ethernet frames at designated times, as part of their 5T for 5G feature set. They also provide hardware timestamping for received and transmitted frames. Although these functions are expected to satisfy the stringent timing requirements of 5G fronthaul, little public information is available regarding their timing accuracy and performance characteristics.
This paper presents an experimental characterization of the Accurate Scheduling and hardware timestamping capabilities of the NVIDIA ConnectX-7 NIC.
Using an FPGA-based measurement platform with deterministic frame generation and nanosecond-resolution timestamping, we first evaluate the precision of the receive and transmit hardware timestamps and then evaluate the transmission timing accuracy of Accurate Scheduling.
The experimental results show that the receive and transmit hardware timestamps exhibit a measured variation of approximately $\pm$7-8 ns when compared across independently clocked Ethernet entities.
Furthermore, Accurate Scheduling transmits approximately 99\% of frames within $\pm$900 ns of the specified transmission time, while occasional outliers of up to approximately 5 us are observed. These results indicate that Accurate Scheduling is well suited for applications with latency requirements on the order of several tens of microseconds, such as 5G fronthaul, whereas its timing accuracy is insufficient for highly deterministic TSN applications, which typically require transmission timing accuracy on the order of several to several tens of nanoseconds.
\end{abstract}

\begin{IEEEkeywords}
NIC, scheduled transmission, hardware timestamp, Ethernet, ConnectX, Accurate Scheduling.
\end{IEEEkeywords}

\section{Introduction}

High-precision packet transmission is becoming increasingly important in modern networks. In the 5G fronthaul of C-RAN (Cloud Radio Access Network), distributed units (DUs) and radio units (RUs) need to exchange data at precisely controlled times to satisfy stringent timing requirements\cite{ecprireq} defined by eCPRI\cite{ecpri}. Recent NVIDIA ConnectX NICs support Accurate Scheduling, a hardware-assisted mechanism that enables Ethernet frames to be transmitted at designated times.
This functionality is part of NVIDIA's 5T for 5G (Time-Triggered Transmission Technology for Telco), which also includes precise hardware timestamping of transmitted and received frames.
Hardware timestamping is indispensable for building time-synchronized communication systems\cite{1588ptp}.

Beyond its original target application, such a hardware-assisted transmission mechanism could also be useful for industrial Ethernet networks\cite{P60802,8021DG,P8021DP} implementing the transmission mechanisms defined by Time-Sensitive Networking (TSN), such as Scheduled Traffic (also known as Time-aware Shaping, TAS) and Asynchronous Traffic Shaping (ATS) \cite{ieee8021q2022}.

Since Accurate Scheduling is designed for 5G fronthaul, it is expected to provide sufficiently accurate transmission timing for such applications. The eCPRI specification defines stringent latency requirements for fronthaul networks, including a one-way frame delay of no more than 25 us for ultra-low-latency use cases \cite{ecprireq}. However, little publicly available information exists to quantitatively characterize the timing accuracy of Accurate Scheduling or to determine whether it satisfies the requirements of these demanding applications. Consequently, its applicability to other deterministic networking applications, including TSN, also remains unclear.
Likewise, the accuracy characteristics of the hardware timestamping functionality have not been sufficiently characterized in the public literature.

This paper presents an experimental characterization of Accurate Scheduling and hardware timestamping on NVIDIA ConnectX-6 Dx and ConnectX-7 NICs.
While ConnectX-6 Dx implements one mode of Accurate Scheduling using an existing synchronization mechanism between multiple queue pairs, ConnectX-7 additionally supports another mode that directly specifies the transmission time in each transmit descriptor. Since ConnectX-7 supports both modes, this paper primarily compares their timing characteristics.
To enable precise evaluation of transmission timing and hardware timestamping, we employ an FPGA-based frame generator and capture platform that provides deterministic frame generation and nanosecond-resolution measurements.
We first characterize the receive and transmit hardware timestamping functionalities to gain insights into the underlying timing mechanisms implemented in the NIC, and then evaluate the timing accuracy of the two Accurate Scheduling modes.
The remainder of this paper is organized as follows. Section II provides an overview of Accurate Scheduling and describes its two scheduling modes. Section III evaluates the accuracy of the receive and transmit hardware timestamps. Section IV evaluates the transmission timing accuracy of Accurate Scheduling. Section V describes related work. Finally, Section VI concludes the paper.

\section{Overview of Accurate Scheduling}

At the time of writing, Linux does not provide native support for the Accurate Scheduling functionality of NVIDIA ConnectX NICs. In contrast, the Data Plane Development Kit (DPDK) has supported this functionality since July 2020 for ConnectX-6 Dx and since February 2022 for ConnectX-7.

Although NVIDIA publicly introduces Accurate Scheduling, no publicly available technical document provides sufficient information on its implementation, operating mechanism, or timing accuracy. Therefore, to provide the background necessary for the subsequent evaluation, we analyzed the DPDK implementation and its development history, including the corresponding source code and commit history.
For convenience, this paper refers to the two implementations as the Clock Queue mode (used in ConnectX-6 Dx) and the Send Queue mode (introduced in ConnectX-7). The following subsections summarize their operation.

\subsection{The Clock Queue mode}

In the Clock Queue mode, which is used by ConnectX-6 Dx and also supported by ConnectX-7, the transmission time cannot be specified directly in the WQEs of a Send Queue (SQ). Instead, scheduled transmission is implemented by leveraging the synchronization mechanism between queue pairs.

The software first creates a dedicated, pseudo Send Queue, referred to as the Clock Queue, which is not used for actual frame transmission. Instead of frame transmission requests, the Clock Queue contains WQEs with NOP (No Operation) commands. The software configures the execution interval ({\tt tx\_pp}) of the Clock Queue so that these NOP WQEs are processed periodically by the NIC hardware.

Whenever a NOP WQE is executed, the hardware posts a completion entry to the associated Completion Queue (CQ). Consequently, the Producer Index of the CQ advances at a constant interval synchronized with the NIC hardware clock.

To transmit a frame at a designated time, the software inserts a WAIT WQE immediately before the corresponding transmit WQE in the normal Send Queue. The WAIT WQE specifies the target Producer Index of the Clock Queue's Completion Queue, which is calculated from the desired transmission time. When processing the WAIT WQE, the hardware suspends execution of the Send Queue until the Producer Index reaches the specified value. Once the target Producer Index is reached, processing resumes and the subsequent transmit WQE is executed, resulting in transmission at the scheduled time.

\subsection{The Send Queue mode}

In the Send Queue mode, introduced in ConnectX-7, the transmission time can be specified directly in a WAIT WQE in the Send Queue.
To transmit a frame at a designated time, the software inserts a WAIT WQE with the wait\_on\_time option immediately before the corresponding transmit WQE in the Send Queue. This WAIT WQE directly specifies the target transmission time.

When processing such a WAIT WQE, the NIC hardware suspends execution of the Send Queue until the hardware clock reaches the specified time. Once the target time is reached, execution resumes and the subsequent transmit WQE is processed, resulting in transmission at the scheduled time.

Because the target time is encoded directly in the WAIT WQE, this implementation does not require a dedicated Clock Queue or periodic NOP processing, thereby simplifying the implementation of scheduled transmission.

\section{Evaluation of Receive and Transmit Hardware Timestamping}

NVIDIA ConnectX NICs provide hardware timestamping for transmitted and received frames. The NIC maintains a PTP Hardware Clock (PHC) based on International Atomic Time (TAI), which enables hardware timestamping and supports the Precision Time Protocol (PTP)\cite{1588ptp}. By referencing the PHC, the NIC records the transmission and reception times of frames with high precision.

Accurate Scheduling also relies on the PHC to determine when each frame should be transmitted. Consequently, hardware timestamping and Accurate Scheduling share a common hardware time base and are expected to exhibit similar timing characteristics. Therefore, before evaluating Accurate Scheduling, we first characterize the hardware timestamping functionality to better understand the timing behavior of the underlying hardware.

To accurately evaluate the hardware timestamping accuracy with nanosecond-level precision, we use the open-source FPGA-based Ethernet Frame Crafter and Capture (EFCC) platform\cite{CCIRT2024} developed in our previous work\cite{efcc}. Originally designed for Gigabit Ethernet, EFCC has been extended in this work to support 10GbE. EFCC can generate arbitrary frame sequences according to predefined traffic patterns while controlling the transmission interval with a resolution of 6.4 ns. It also records frame transmission and reception times with the same temporal resolution. In this paper, EFCC is implemented on a Xilinx Alveo U250 FPGA board equipped with two QSFP28 ports.

Note that, throughout this paper, the frame interval is defined as the time from the first bit of a frame to the first bit of the subsequent frame (i.e., the start-to-start interval). The standard maximum Ethernet frame size is 1518 bytes, measured from the destination MAC address field to the Frame Check Sequence (FCS) field. At the physical layer, each frame is preceded by a 7-byte Preamble and a 1-byte Start Frame Delimiter (SFD), followed by a 12-byte Interframe Gap (IFG). Consequently, the back-to-back frame interval for a maximum-sized frame corresponds to the transmission time of 1538 bytes at the line rate. To eliminate variations in the frame interval, the Deficit Idle Count (DIC) mechanism, which dynamically adjusts the length of the IFG while preserving the average IFG required by IEEE 802.3, was not used in EFCC.

In the following experiments, the ConnectX-7 NIC was installed in a PCI Express interface of a host machine equipped with an Intel Xeon w5-3435X processor (16 CPU cores). The evaluated ConnectX-7 adapter was model MCX755106AS-HEAT with firmware version 28.48.1000.
To improve the reproducibility of the experimental results, Simultaneous Multithreading (SMT), Dynamic Voltage and Frequency Scaling (DVFS), and CPU idle states were disabled. PCI Express Active State Power Management (ASPM) was left disabled (the default setting). NVIDIA DOCA SDK version 3.3.0109 was used. The host ran Ubuntu 24.04 LTS with Linux kernel version 6.8.0-110-generic.
For the Accurate Scheduling evaluation, we developed a frame transmission program using DPDK version 25.11.0+doca2601. To eliminate interference from other operating system tasks, a dedicated CPU core was isolated from the Linux scheduler at boot time and exclusively assigned to the transmission program throughout the experiments.

The ConnectX-7 NIC requires a PCIe Gen5 x16 interface. The mainboard of the host machine (Supermicro X13SWA-TF) provides a PCIe Gen5 ×16 slot and therefore satisfies this requirement. We also confirmed that similar experimental results were obtained using another host machine equipped with a different motherboard.

\subsection{Validation of the measurement platform}

Before evaluating the NIC hardware, we validated the accuracy of the EFCC measurement platform. The two QSFP28 ports on the Alveo U250 board were directly connected using a QSFP+ DAC cable. First, 50,000 frames of 1518 bytes were transmitted back-to-back at the 10GbE line rate. The receive timestamps recorded by the opposite port showed an frame interval of exactly 1235.2 ns (193 cycles) for all frames, which matches the theoretical value. This result confirms that EFCC accurately measures frame arrival times. Furthermore, the propagation delay between the two ports remained constant for all transmitted frames, indicating that no measurable timing variation was introduced by the measurement platform.

Next, the transmission interval was configured to 6.4 us (1000 cycles), and another 50,000 frames were transmitted. The measured receive timestamps showed a frame interval of exactly 6.4 us for every frame, confirming that EFCC introduces no measurable transmission interval error.

\subsection{Evaluation of receive hardware timestamping}

To evaluate the accuracy of the receive hardware timestamping functionality, one port of EFCC was directly connected to a port of the ConnectX-7 NIC using a QSFP+ DAC cable. EFCC transmitted 50,000 frames of a fixed size back-to-back at the 10GbE line rate. The ConnectX-7 NIC recorded the receive hardware timestamp of each frame, and the measured frame intervals were compared with the corresponding theoretical values determined by the frame size. Receive hardware timestamps were collected using {\tt tcpdump} with hardware timestamping enabled.

For experiments using the minimum Ethernet frame size (64 bytes), multiple UDP flows were generated by varying the source UDP port number for each transmitted frame in a round-robin manner. This configuration distributes incoming frames across multiple receive queues through the Receive Side Scaling (RSS) mechanism in ConnectX-7, preventing frame drops caused by receive queue overflow.

ConnectX-7 supports two modes for maintaining the NIC's internal clock: the free-running clock and the real-time clock. The free-running clock maintains the elapsed time since NIC initialization, whereas the real-time clock maintains an absolute time reference such as International Atomic Time (TAI). Both clock modes were evaluated in the following experiment.

We investigated the format of receive hardware timestamps recorded in the Completion Queue (CQ) using techniques such as eBPF. In both clock modes, timestamps are represented in nanoseconds. However, the least significant bit (LSB) of the receive hardware timestamp is always set to 1, regardless of whether the free-running clock or the real-time clock is used.
This observation suggests that the LSB is used as a mode indicator rather than as part of the timestamp value. Consequently, the effective timestamp resolution is inferred to be 2 ns for both clock modes.

\begin{table*}[t]
  \centering
  \caption{Distribution of frame intervals derived from ConnectX-7 receive hardware timestamps for back-to-back transmission of 64-byte frames}
  \label{tab:rxts64b}
  \begin{tabular}{c|rrrrrrrrr|r|r}
    \hline
    & \multicolumn{11}{c}{Frame interval (ns)} \\ \hline
    &  62 & 64 & 66 & 68 & 70 & 72 & 74 & 76 & 78 & Mean & Std. \\ \hline
    Free-running clock & 0 & 160 & 2387 & 11346 & 15694 & 14586 & 5474 & 352 & 0  & 70.4 & 2.2 \\
    Real-time clock   & 2 & 104 & 1962 & 11974 & 16315 & 13817 & 5110 & 686 & 29 & 70.4 & 2.2  \\
    \hline
  \end{tabular}
\end{table*}

\begin{table*}[t]
  \centering
  \caption{Distribution of frame intervals derived from ConnectX-7 receive hardware timestamps for back-to-back transmission of 1518-byte frames}
  \label{tab:rxts1518b}
  \begin{tabular}{c|rrrrrrrrr|r|r}
    \hline
    & \multicolumn{11}{c}{Frame interval (ns)} \\ \hline
    &  1228 & 1230 & 1232 & 1234 & 1236 & 1238 & 1240 & 1242 & 1244 & Mean & Std. \\ \hline
    Free-running clock &    0 & 986 & 7437 & 14897 & 16202 & 8573 & 1703 & 201 & 0 & 1235.2   & 2.2  \\
    Real-time clock   &   52 & 1241 & 7542 & 14901 & 15031 & 8904 & 2103 & 221 & 4 &   1235.2   & 2.3  \\
    \hline
  \end{tabular}
\end{table*}

The results for the 64-byte and 1518-byte frame sizes are summarized in Tables \ref{tab:rxts64b} and \ref{tab:rxts1518b}, respectively. At the physical layer, a 64-byte frame corresponds to 84 bytes on the wire and requires 11 clock cycles (70.4 ns) for transmission in 10GbE. Similarly, a 1518-byte frame requires 193 clock cycles (1235.2 ns). In all experiments, the mean frame interval derived from the receive hardware timestamps exactly matched the corresponding theoretical value. Furthermore, the measured frame intervals exhibited a symmetric distribution centered around the theoretical value. The highest occurrence was observed at 70 ns and 72 ns for the 64-byte frame, and at 1234 ns and 1236 ns for the 1518-byte frame. This distribution is consistent with the inferred 2-ns timestamp resolution discussed above.

For the free-running clock mode, the difference between the minimum and maximum measured frame intervals was 12 ns for both frame sizes. Since the timestamp may vary by approximately one clock cycle (6.4 ns) due to clock-domain crossing, the observed 12-ns variation in the frame interval can be explained by the accumulation of timestamp deviations at both the preceding and succeeding frames. Furthermore, the distributions obtained from the 1518-byte experiment using a single flow and the 64-byte experiment using multiple flows exhibited nearly identical characteristics. This result indicates that the Receive Side Scaling (RSS) mechanism does not have a measurable impact on the accuracy of receive hardware timestamping.

For the real-time clock mode, the difference between the minimum and maximum measured frame intervals was 16 ns for both frame sizes, which is 4 ns larger than that observed for the free-running clock mode. In the free-running clock mode, the timestamp stored in the Completion Queue (CQ) entry represents the elapsed time since NIC initialization and is subsequently converted into an absolute timestamp by the device driver. In contrast, in the real-time clock mode, the timestamp recorded in the CQ entry is already represented in the PHC time domain and therefore requires no further conversion by the device driver. The additional variation observed in the real-time clock mode suggests that the timestamp generation path within the NIC is slightly more complex than that of the free-running clock mode, although the underlying implementation is not publicly documented.

\subsection{Evaluation of transmit hardware timestamping}

The ConnectX-7 NIC and EFCC remained directly connected as in the previous experiment. Using the receive hardware timestamps recorded by EFCC as a reference, we evaluated the accuracy of the transmit hardware timestamping functionality of ConnectX-7.

A frame transmission program was executed on the host connected to the ConnectX-7 NIC. The Linux socket option for transmit hardware timestamping was enabled, and 10,000 frames of 1518 bytes were transmitted at a constant interval under software control. After transmission, the hardware transmit timestamp of each frame was retrieved from the socket error queue.

In Linux, hardware transmit timestamps are delivered asynchronously through the socket error queue. The maximum number of timestamps that can be buffered depends on the receive buffer size of the socket ({\tt SO\_RCVBUF}). With the default configuration (208 KiB), timestamps could be retrieved for only approximately 290 transmitted frames before subsequent timestamps were discarded. Therefore, the receive buffer size was increased to 128 MiB in the experiment.

To evaluate the accuracy of transmit hardware timestamping independently of the Accurate Scheduling functionality, frame transmission timing was controlled solely by software. Consequently, the actual transmission times were not perfectly periodic and could exhibit sub-millisecond timing variations due to software scheduling. If the transmit hardware timestamping functions correctly, however, it should accurately record the actual transmission time of each frame regardless of when the frame is transmitted.

For transmit hardware timestamping, the tx\_port\_ts option was enabled so that timestamps were recorded at the instant the frame was transmitted from the NIC port, rather than when the corresponding Completion Queue (CQ) entry was generated.
Since Accurate Scheduling requires the real-time clock mode, the following experiments focus on the results obtained with this mode. We also confirmed that the results obtained with the free-running clock mode exhibit the same characteristics.

In contrast to receive hardware timestamps, whose least significant bit (LSB) is always set to 1, the LSB of transmit hardware timestamps is always set to 0. As with receive hardware timestamps, transmit hardware timestamps are represented in nanoseconds in both clock modes. The effective timestamp resolution is therefore inferred to be 2 ns.

Since the ConnectX-7 NIC and EFCC are directly connected through a QSFP+ DAC cable, the propagation delay between them is essentially constant, except for a small variation of at most a few clock cycles caused by factors such as clock-domain crossings within the devices. Therefore, after compensating for the clock drift between the two devices, the variation in the difference between the transmit hardware timestamps reported by ConnectX-7 and the receive hardware timestamps recorded by EFCC provides an estimate of the accuracy of the transmit hardware timestamping functionality.

\begin{figure}[!t]
\centering
\includegraphics[width=\columnwidth]{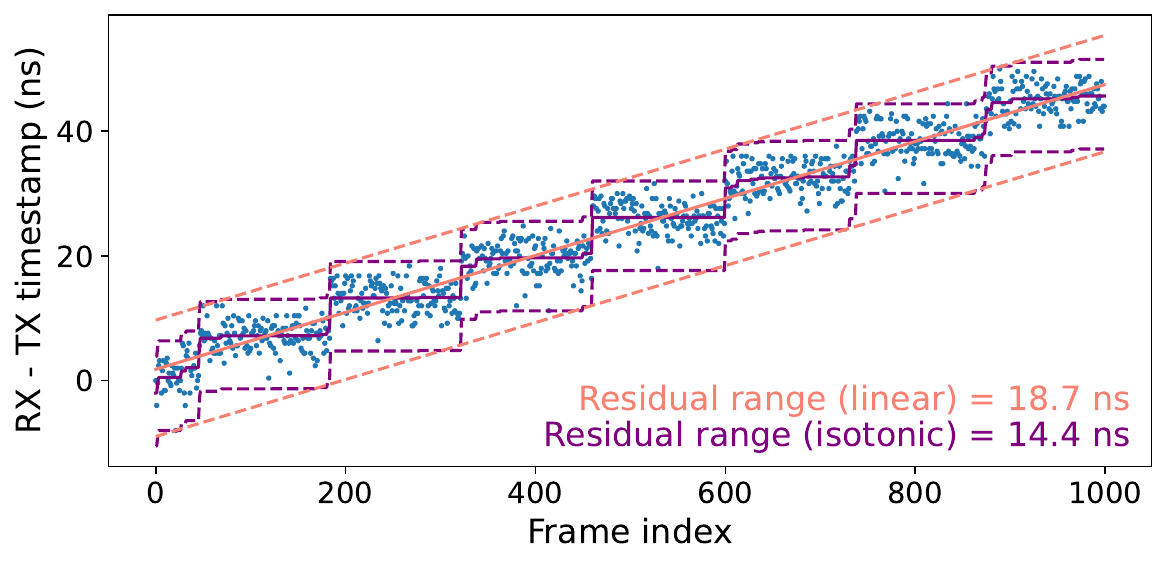}
\caption{Relative difference between the ConnectX-7 transmit hardware timestamps and the reference receive hardware timestamps, normalized to the first frame. The solid lines show the linear (red) and isotonic (purple) regressions, and the corresponding dashed lines indicate their residual ranges.}
\label{fig:txts}
\end{figure}

The frame transmission interval was set to 10 us. Figure\ref{fig:txts} shows the difference between the transmit hardware timestamp reported by ConnectX-7 and the receive hardware timestamp recorded by EFCC, normalized to the value of the first frame. For clarity, only the first 1,000 frames of the 10,000 transmitted frames are shown, although the same behavior was observed throughout the entire experiment. 
The solid lines show the linear (red) and isotonic (purple) regressions, and the corresponding dashed lines indicate their residual ranges.

The apparent propagation delay relative to that of the first frame increased approximately linearly over time. This trend was caused by the clock drift between the ConnectX-7 NIC and EFCC. Specifically, the clock of the transmitting side (ConnectX-7) ran slightly slower than that of the receiving side (EFCC). For example, for the 10,000th frame, which was transmitted approximately 100 ms after the first frame, the transmit-side clock lagged behind the receive-side clock by approximately 470 ns. Assuming that the oscillator frequency offset of each device remains constant under stable temperature conditions, the relative clock drift can be accurately modeled and compensated for using a linear function.

A closer inspection reveals that the apparent propagation delay exhibits a staircase-like distribution. This staircase pattern becomes more evident when the data are fitted using isotonic regression. The observed step intervals are consistent with the 6.4 ns resolution of the EFCC receive hardware timestamp, which records packet arrival times once every clock cycle.

The upper and lower bounds of the residuals from the linear and isotonic regression are also shown in Figure \ref{fig:txts}. The residual ranges of the linear and isotonic regressions were 18.7 ns and 14.4 ns, respectively. The latter value indicates that the transmit hardware timestamping accuracy of ConnectX-7 is approximately $\pm$7-8 ns. This result is comparable to the 16-ns variation observed in the frame intervals derived from the receive hardware timestamps when the real-time clock mode was used. The similarity between these values suggests that the timestamp generation mechanisms for transmit and receive operations may share a common implementation within the NIC.
The same behavior was observed when the frame transmission interval was increased to 100 us and 1 ms.

In contrast, when frames were transmitted back-to-back without an intentional interval, the transmit hardware timestamps were not always correctly recorded in the Completion Queue (CQ). In some cases, multiple consecutive frames were assigned identical transmit timestamps. By comparing these timestamps with the receive hardware timestamps recorded by EFCC, we found that timestamp recording failures occurred when the actual frame interval became shorter than approximately 1-2 us.

Further analysis revealed that this behavior occurred when multiple transmit descriptors were submitted to the Send Queue with a single doorbell update. In such cases, the corresponding frames were assigned the same transmit hardware timestamp.
This observation suggests that the generation of a CQE containing a hardware timestamp is associated with a doorbell update rather than with individual frame transmissions when multiple descriptors are posted simultaneously.

According to NVIDIA's publicly available documentation, the transmit hardware timestamping functionality is primarily intended for PTP applications. Since PTP synchronization messages are typically transmitted at relatively low rates rather than back-to-back, the hardware timestamping implementation may not have been designed to support continuous timestamp generation for high-rate packet transmissions.

\subsection*{\bf Summary:}
The experimental results show that the receive and transmit hardware timestamps exhibit a measured variation of approximately $\pm$7-8 ns. This variation includes the inherent uncertainty associated with comparing timestamps across independently clocked Ethernet entities and therefore should be regarded as an upper bound on the intrinsic timestamping error of the ConnectX-7 NIC.

\section{Evaluation of Accurate Scheduling}

Using the same experimental setup as in the previous experiment, we evaluated the timing accuracy of Accurate Scheduling on ConnectX-7 by comparing the receive hardware timestamps recorded by EFCC with the scheduled transmission times. A frame transmission program was implemented using DPDK and executed on the host connected to the ConnectX-7 NIC. The program transmitted 50,000 frames, each associated with a designated transmission time. The transmission time of each frame was specified such that consecutive frames were scheduled at 100-us intervals. Experiments were conducted using both 64-byte and 1518-byte frames. As described above, the NIC was configured to operate in the real-time clock mode, which is required for Accurate Scheduling.
For the Clock Queue mode, the NOP execution interval parameter (\texttt{tx\_pp}) was set to either the minimum supported interval (500 ns) or 5 us. For comparison, the Send Queue mode was also evaluated.

\begin{table}[t]
  \centering
  \caption{Distribution of actual frame intervals for 64-byte frames}
  \label{tab:as64b}
  \begin{tabular}{c|rr|r}
    \hline
                  & \multicolumn{2}{c|}{Clock Queue Mode} & \multicolumn{1}{c}{Send Queue Mode} \\
                  & tx\_pp = 500  & 5,000            &  \\ \hline
    Min (ns)      &  97,184 & 	 98,086 & 	96,332 \\
    $P_{0.5}$ (ns) & 99,277 &    99,181  &   99,270 \\
    $P_{99.5}$ (ns) & 100,717 &   100,826 &  100,742 \\
    Max (ns)      & 102,188 &	101,203 & 101,113  \\ \hline
    Max - Min (ns) &  4,928 &	  4,026 &	 5,780   \\
    Mean (ns)     & 100,000.0 	& 100,000.0  & 	99,999.9  \\
    Std. (ns)     & 199.3	& 218.1	& 205.8  \\
    \hline
  \end{tabular}
\end{table}

\begin{table}[t]
  \centering
  \caption{Distribution of actual frame intervals for 1518-byte frames}
  \label{tab:as1518b}
  \begin{tabular}{c|rr|r}
    \hline
                  & \multicolumn{2}{c|}{Clock Queue Mode} & \multicolumn{1}{c}{Send Queue Mode} \\
                  & tx\_pp = 500  & 5,000            &  \\ \hline
    Min (ns)      & 95,750 & 95,936	 & 96,518 \\
    $P_{0.5}$ (ns) & 99,264 &  99,168 & 99,123 \\
    $P_{99.5}$ (ns) & 100,736  & 100,825  & 100,890 \\
    Max (ns)      & 102,112	& 101,766 & 103,513  \\ \hline
    Max - Min (ns) & 6,362	& 6,176 & 	5,594  \\
    Mean (ns)     & 99,999.9 & 99,999.9 & 99,999.9 \\
    Std. (ns)     & 211.9 & 	228.1	& 245.2 \\
    \hline
  \end{tabular}
\end{table}

Tables \ref{tab:as64b} and \ref{tab:as1518b} summarize the measured frame intervals observed by EFCC for 64-byte and 1518-byte frames, respectively. In all configurations, the mean frame interval was 100 us or very close to it, indicating that the target transmission interval specified by the transmission program was maintained on average. On the other hand, the standard deviation ranged from approximately 200 to 250 ns, indicating that the actual transmission intervals exhibited a certain amount of jitter. In addition, the difference between the maximum and minimum observed frame intervals ranged from 4.0 to 6.4 us. Despite this jitter, the observed timing variation was substantially smaller than that of software-based transmission, which typically exhibits timing variations on the order of sub-milliseconds.

One might expect that a smaller value of the Clock Queue execution interval parameter (\texttt{tx\_pp}) would improve the scheduling accuracy by allowing transmission timing to be controlled more frequently. However, no systematic difference in the distribution of the actual frame intervals was observed between \texttt{tx\_pp} = 500 ns and 5 us. Similarly, the Send Queue mode, which directly specifies transmission timestamps in the transmit descriptors without using the Clock Queue, was expected to achieve higher scheduling accuracy than the Clock Queue mode. However, no systematic difference attributable to the scheduling interface mode was observed.

\begin{figure}[!t]
\centering
\includegraphics[width=\columnwidth]{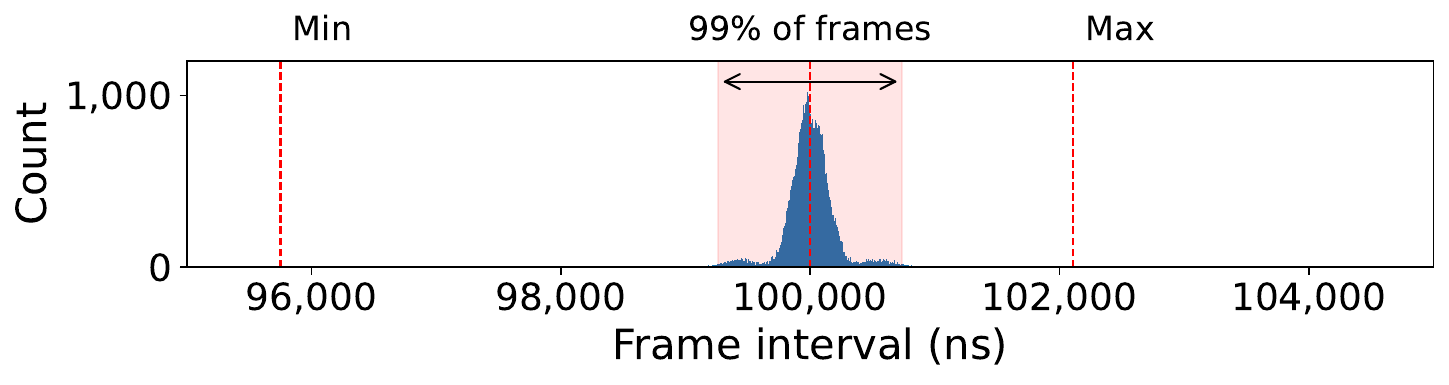}
\caption{Distribution of actual frame intervals for 1518-byte frames in the Clock Queue mode (\texttt{tx\_pp} = 500 ns)}
\label{fig:ascq5001538}
\end{figure}

\begin{figure}[!t]
\centering
\includegraphics[width=\columnwidth]{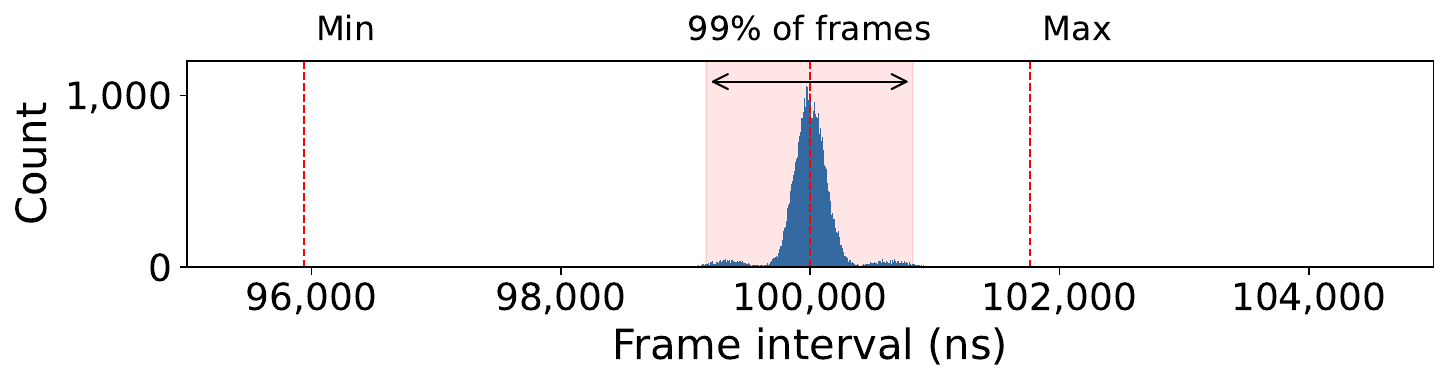}
\caption{Distribution of actual frame intervals for 1518-byte frames in the Clock Queue mode (\texttt{tx\_pp} = 5000 ns)}
\label{fig:ascq50001538}
\end{figure}

\begin{figure}[!t]
\centering
\includegraphics[width=\columnwidth]{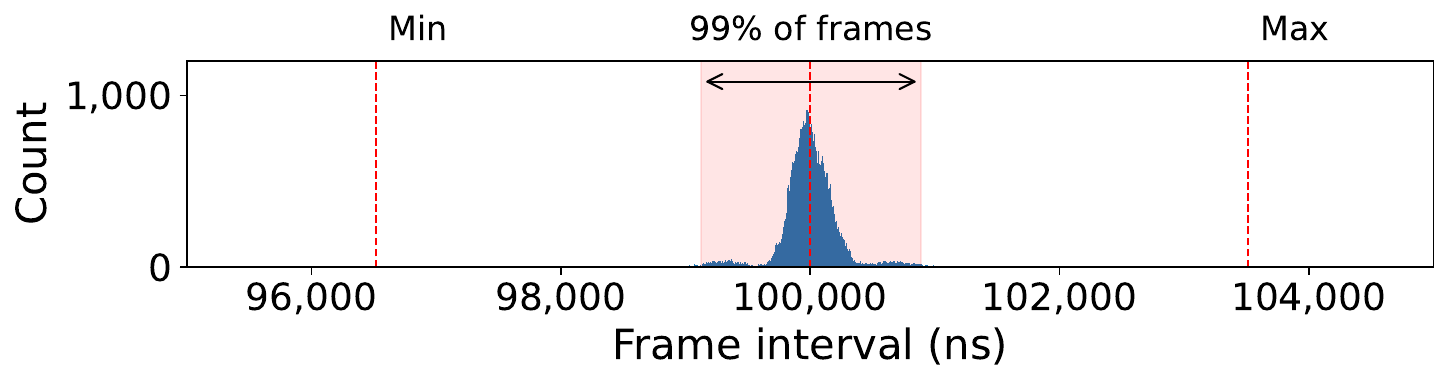}
\caption{Distribution of actual frame intervals for 1518-byte frames in the Send Queue mode}
\label{fig:assq1538}
\end{figure}

Figures \ref{fig:ascq5001538}--\ref{fig:assq1538} show histograms of the actual frame intervals. Since no significant differences were observed between the results for 64-byte and 1518-byte frames, only the results for 1518-byte frames, which are representative of a wide range of application traffic, including typical eCPRI user-plane traffic, are presented. The histogram bin width was set to 6.4 ns, corresponding to the timestamp resolution of the EFCC receive hardware timestamps. In all configurations, approximately 99\% of the measured frame intervals were within $\pm$900 ns of the target interval of 100 us.
Upon closer inspection, small secondary peaks were observed around 99.5 and 100.5 us. Similar peaks with an interval of approximately 500 ns were consistently observed for different frame sizes and target transmission intervals.

\begin{figure}[!t]
\centering
\includegraphics[width=\columnwidth]{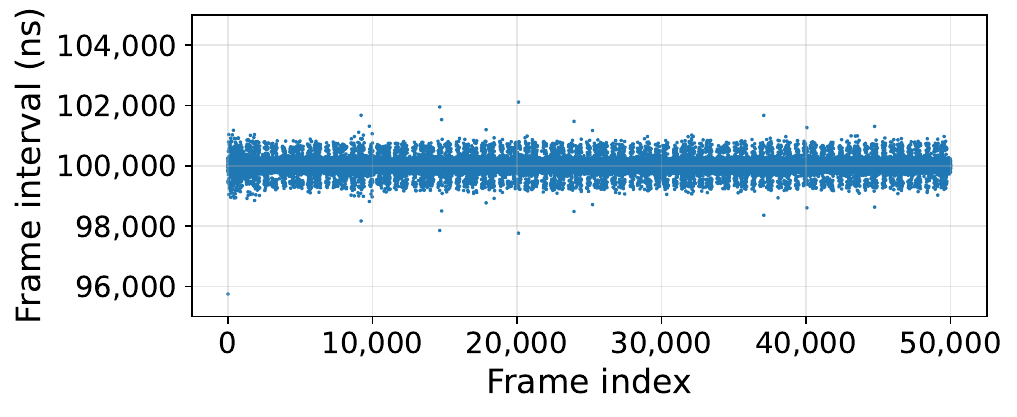}
\caption{Actual frame intervals versus frame index for 1518-byte frames in the Clock Queue mode (\texttt{tx\_pp} = 500 ns)}
\label{fig:ascq5001538ts}
\end{figure}

\begin{figure}[!t]
\centering
\includegraphics[width=\columnwidth]{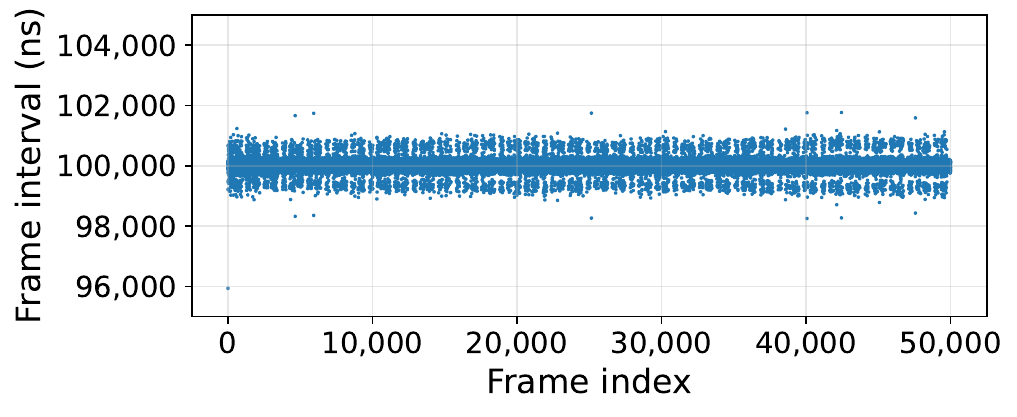}
\caption{Actual frame intervals versus frame index for 1518-byte frames in the Clock Queue mode (\texttt{tx\_pp} = 5000 ns)}
\label{fig:ascq50001538ts}
\end{figure}

\begin{figure}[!t]
\centering
\includegraphics[width=\columnwidth]{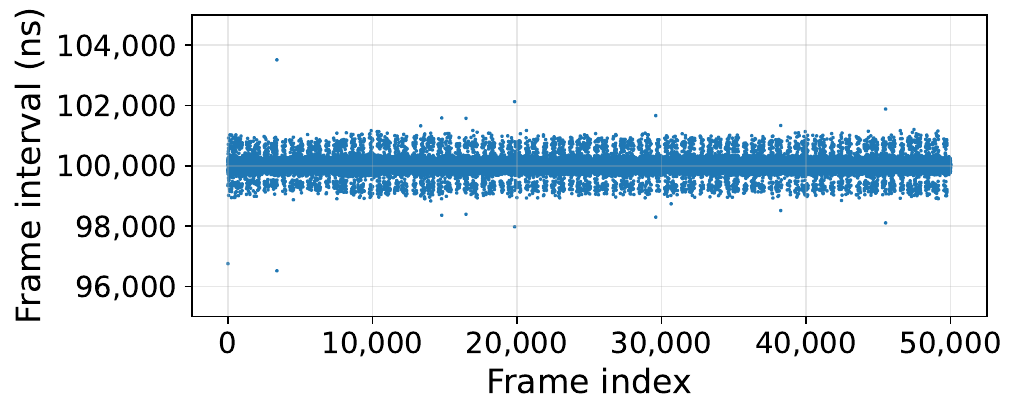}
\caption{Actual frame intervals versus frame index for 1518-byte frames in the Send Queue mode}
\label{fig:assq1538ts}
\end{figure}

Figures \ref{fig:ascq5001538ts}--\ref{fig:assq1538ts} show the actual frame intervals plotted against the frame index. Consistent with the histogram results, the majority of the frame intervals remained within approximately $\pm$900 ns of the target interval. However, the spread of the frame intervals varied over time with an apparent periodic trend. In addition, a small number of outliers were observed, where the frame interval deviated from the target value by several microseconds.
Neither the histogram nor the time-series plots revealed any systematic differences attributable to the scheduling interface mode or the value of \texttt{tx\_pp}.

\subsection*{\bf Summary:}

Overall, the experimental results demonstrate that Accurate Scheduling can transmit approximately 99\% of the frames within $\pm$900 ns of the specified transmission time. Although outliers with deviations of several microseconds, up to approximately 5 us, were observed, they occurred only rarely. This level of timing accuracy is difficult to achieve using software-based transmission alone and is expected to be sufficient for applications with latency requirements on the order of several tens of microseconds, such as ultra-low-latency eCPRI.
In contrast, the achieved accuracy is insufficient for 10GbE TSN applications, such as TAS and ATS, where timing accuracy on the order of several to several tens of nanoseconds is generally required. Although the exact requirement depends on the application and network configuration, the transmission-time error should be kept sufficiently smaller than the transmission time of a minimum-sized frame (67 ns at 10GbE) to avoid effects such as frame reordering and increased deviation from the worst-case delay bound.

The cause of the outliers remains unclear. Since the transmission program exclusively occupied a dedicated CPU core throughout the experiments, software-related factors, such as delayed doorbell notifications caused by interrupt handling, are unlikely to be the primary cause. Instead, the outliers are more likely to originate from processing within the NIC hardware.

No systematic differences attributable to either the scheduling interface mode (i.e., the Clock Queue mode or the Send Queue mode) or the value of the Clock Queue execution interval parameter (\texttt{tx\_pp}) were observed. This suggests that the observed transmission jitter is unlikely to originate from the software--NIC interface, but rather from the internal transmission scheduling mechanism within the NIC. Moreover, the secondary peaks observed at approximately 500 ns intervals suggest the presence of an internal scheduling mechanism operating at a similar time granularity.

This interpretation is further supported by the fact that the previously evaluated transmit and receive hardware timestamping accuracy of ConnectX-7 was no more than approximately $\pm$7-8 ns, which is substantially smaller than the transmission timing jitter observed in this evaluation. This suggests that the dominant source of the observed jitter is more likely to reside in the transmission timing control mechanism than in the NIC timekeeping logic.

Finally, the transmission timing accuracy exhibited an apparent periodic variation over time, alternating between periods of relatively small and relatively large jitter. Such behavior could potentially result from periodic changes in the operating frequency caused by power management or thermal protection mechanisms. However, no evidence supporting these possibilities was found while monitoring the operating status of the NIC.

\section{Related Work}

Similar to NVIDIA Accurate Scheduling, several commercial Ethernet controllers provide hardware-assisted transmission scheduling capabilities for deterministic networking applications\cite{P60802,8021DG,P8021DP}. Intel Ethernet controllers such as the I210 and I225 support LaunchTime, which allows software to specify the transmission time of each frame. This functionality is intended to support software implementations of the Scheduled Traffic mechanism (commonly referred to as Time-Aware Shaping, TAS) defined in IEEE 802.1Q-2022. STMicroelectronics also provides a similar functionality called Time-Based Scheduling (TBS) in some of their Ethernet controller products. However, these controllers are limited to link speeds of up to 1 or 2.5 GbE, making them unsuitable for applications requiring 10GbE or higher, such as 5G fronthaul and low-latency user-plane processing nearby a 5G core network\cite{mec003,mec002}.

The Intel I210 and I225 are widely used in off-the-shelf GbE NICs. In particular, the I210 has been used in several TSN studies because its programming interface is publicly documented\cite{i210datasheet}, and a number of studies have reported performance evaluations of its LaunchTime functionality\cite{method2022,oge2020,dac2024}.

In our previous work\cite{access2025hirofuchi}, we developed an ATS endpoint based on the Intel I210 and conducted a more rigorous evaluation of its transmission timing accuracy, particularly under high-rate transmission conditions, to determine whether it could satisfy the timing requirements of ATS. We confirmed that the I210 is capable of transmitting frames at specified transmission times with an accuracy of approximately 48 ns under suitable conditions.

However, we also found that its timing accuracy depends on both the PCIe implementation of a transmitting host and frame size. In one host machine, although its PCIe interface satisfied the requirements of the I210, for example, the transmission timing accuracy for 1518-byte frames degraded rapidly when the frame interval became shorter than approximately 18 us. As the transmission interval decreased, the timing error increased to nearly 10 us. The NIC was unable to sustain the target transmission rate, resulting in frame drops.

According to the Intel I210 datasheet\cite{i210datasheet}, enabling LaunchTime changes the way the NIC fetches frame data from host memory. The datasheet also states that the controller provides a 24-KB on-chip transmit buffer.
Since the on-chip buffer is shared among multiple hardware transmit queues, using the entire 24 KB buffer to prefetch frames for a single hardware queue would prevent the NIC from transmitting frames from other hardware queues when they become eligible for transmission. Such behavior would violate the scheduling semantics required by IEEE802.1Q, in which the interactions among different traffic classes are strictly specified.
A plausible explanation is therefore that, when LaunchTime is enabled, the NIC limits the amount of frame data prefetched from host memory for each hardware queue. The transmission process becomes more sensitive to PCIe access latency, which resulted in time errors with a particular PCIe slot.

In contrast, the experiments presented in this paper evaluated Accurate Scheduling on the ConnectX-7 using a transmission interval of 100 us, which is substantially longer than those used in our previous evaluation of the Intel I210.
The transmission timing deviations of up to approximately 5 us observed in this study are unlikely to originate from delays in fetching frame data from host memory over the PCIe bus.

\section{Conclusion}

This paper presented an experimental characterization of the Accurate Scheduling and hardware timestamping capabilities of the NVIDIA ConnectX-7 NIC using an FPGA-based measurement platform with deterministic frame generation and nanosecond-resolution timestamping. 
The experimental results showed that the receive and transmit hardware timestamps exhibit a measured variation of approximately $\pm$7-8 ns, including the inherent uncertainty introduced by comparing timestamps across independently clocked Ethernet devices.
In addition, Accurate Scheduling transmitted approximately 99\% of frames within $\pm$900 ns of the specified transmission time, although occasional outliers of up to approximately 5 us were observed.

The evaluation also provided several insights into the implementation characteristics of Accurate Scheduling. No systematic differences were observed between the Clock Queue and Send Queue modes or between different values of the Clock Queue execution interval parameter. The observed transmission timing jitter was significantly larger than the hardware timestamping error.
This suggests that the dominant source of the observed timing variation does not originate from the scheduling interface or the common timekeeping logic within the NIC. Instead, it is more likely to reside in the transmission timing control mechanism (e.g., gate control) in the NIC hardware.

These results indicate that Accurate Scheduling is well suited for applications with latency requirements on the order of several tens of microseconds, such as 5G fronthaul. However, its timing accuracy remains insufficient for highly deterministic TSN applications that require transmission timing accuracy on the order of several to several tens of nanoseconds. As future work, we plan to investigate whether Accurate Scheduling can be applied to TSN by extending deterministic network models to incorporate bounded transmission timing errors. For example, our ATS endpoint mechanism has a configurable safety margin for transmission timing errors while remaining fully compliant with the ATS state machine defined in the IEEE standard. Although further investigation is required, this mechanism is expected to be applicable to the timing errors of Accurate Scheduling.

\section*{Acknowledgments}
This paper is based on results obtained from the project, "Research and Development Project of the Enhanced infrastructures for Post-5G Information and Communication Systems" (JPNP20017), commissioned by the New Energy and Industrial Technology Development Organization (NEDO).

\bibliographystyle{IEEEtran}
\bibliography{myrefs}

\begin{thebibliography}{10}
\providecommand{\url}[1]{#1}
\csname url@samestyle\endcsname
\providecommand{\newblock}{\relax}
\providecommand{\bibinfo}[2]{#2}
\providecommand{\BIBentrySTDinterwordspacing}{\spaceskip=0pt\relax}
\providecommand{\BIBentryALTinterwordstretchfactor}{4}
\providecommand{\BIBentryALTinterwordspacing}{\spaceskip=\fontdimen2\font plus
\BIBentryALTinterwordstretchfactor\fontdimen3\font minus
  \fontdimen4\font\relax}
\providecommand{\BIBforeignlanguage}[2]{{%
\expandafter\ifx\csname l@#1\endcsname\relax
\typeout{** WARNING: IEEEtran.bst: No hyphenation pattern has been}%
\typeout{** loaded for the language `#1'. Using the pattern for}%
\typeout{** the default language instead.}%
\else
\language=\csname l@#1\endcsname
\fi
#2}}
\providecommand{\BIBdecl}{\relax}
\BIBdecl

\bibitem{ecprireq}
\emph{Common Public Radio Interface: Requirements for the {eCPRI} Transport
  Network}, Common Public Radio Interface, 2018.

\bibitem{ecpri}
\emph{Common Public Radio Interface: {eCPRI} Interface Specification}, Common
  Public Radio Interface, 2019.

\bibitem{1588ptp}
{IEEE Instrumentation and Measurement Society}, \emph{IEEE Standard for a
  Precision Clock Synchronization Protocol for Networked Measurement and
  Control Systems}, Institute of Electrical and Electronics Engineers, 2019.

\bibitem{P60802}
\emph{IEC/IEEE Draft International Standard Time-Sensitive Networking Profile
  for Industrial Automation}, Institute of Electrical and Electronics
  Engineers, 2023.

\bibitem{8021DG}
\emph{IEEE Standard for Local and metropolitan area networks - Time-Sensitive
  Networking Profile for Automotive In-Vehicle Ethernet Communications},
  Institute of Electrical and Electronics Engineers, 2025.

\bibitem{P8021DP}
\emph{Draft Standard for Local and Metropolitan Area Networks: Time-Sensitive
  Networking for Aerospace Onboard Ethernet Communications}, Institute of
  Electrical and Electronics Engineers, 2025.

\bibitem{ieee8021q2022}
\emph{{IEEE} Standard for Local and Metropolitan Area Networks--Bridges and
  Bridged Networks}, Institute of Electrical and Electronics Engineers, 2022.

\bibitem{CCIRT2024}
\emph{The {AIST-TSN} project repository, https://github.com/CCIRT/aist-tsn/},
  2024-2026.

\bibitem{efcc}
A.~B. Ahmed, T.~Hirofuchi, and T.~Fukai, ``Efcc: Ethernet frame crafter and
  capture for tsn research,'' in \emph{Proceedings of the 50th IEEE Conference
  on Local Computer Networks (LCN)}.\hskip 1em plus 0.5em minus 0.4em\relax
  IEEE, 2025, pp. 1--9.

\bibitem{mec003}
\emph{Multi-access Edge Computing (MEC); Framework and Reference Architecture},
  European Telecommunications Standards Institute, Jun 2025.

\bibitem{mec002}
\emph{Multi-access Edge Computing (MEC); Use Cases and Requirements}, European
  Telecommunications Standards Institute, Jun 2025.

\bibitem{i210datasheet}
{Intel Cooperation}, \emph{Intel Ethernet Controller I210 Datasheet, Revision
  Number: 3.7}, 2021.

\bibitem{method2022}
M.~Bosk, F.~Rezabek, K.~Holzinger, A.~G. Marino, A.~A. Kane, F.~Fons, J.~Ott,
  and G.~Carle, ``Methodology and infrastructure for tsn-based reproducible
  network experiments,'' \emph{IEEE Access}, vol.~10, pp. 109\,203--109\,239,
  2022.

\bibitem{oge2020}
Y.~Oge, Y.~Kobayashi, T.~Yamaura, and T.~Maegawa, ``Software-based time-aware
  shaper for time-sensitive networks,'' \emph{IEICE Transactions on
  Communications}, vol. E103.B, no.~3, pp. 167--180, 2020.

\bibitem{dac2024}
C.~Xue, T.~Zhang, and S.~Han, ``Towards cost-effective real-time
  high-throughput end station design for time-sensitive networking (tsn),'' in
  \emph{Proceedings of the 61st ACM/IEEE Design Automation Conference}, ser.
  DAC '24.\hskip 1em plus 0.5em minus 0.4em\relax Association for Computing
  Machinery, 2024.

\bibitem{access2025hirofuchi}
T.~Hirofuchi, A.~B. Ahmed, and T.~Fukai, ``Implementation and evaluation of a
  time-sensitive networking endpoint for asynchronous traffic shaping,''
  \emph{IEEE Access}, vol.~14, pp. 21\,119--21\,136, 2026.

\end{thebibliography}

\end{document}